\documentclass[twocolumn,prb,letterpaper,preprintnumbers,showpacs,floatfix,amsmath,amssymb,superscriptaddress]{revtex4-1}

\usepackage[pdftex]{graphicx}
\usepackage{dcolumn}
\usepackage{bm}
\usepackage{float}
\usepackage{epsfig}
\usepackage{multirow}
\usepackage{amsmath}
\usepackage{amsfonts}
\usepackage{amssymb}
\usepackage{tabularx}

\begin{document}
\title{Theoretical investigation into the possibility of very large
moments in Fe$_{16}$N$_2$}
\author{H. Sims}
\author{W. H. Butler}
\affiliation{Center for Materials for Information Technology (MINT) and Department of Physics, University of Alabama, Tuscaloosa, AL 35487}
\author{M. Richter}
\author{K. Koepernik}
\affiliation{IFW Dresden e.V., P.O. Box 270116, D-01171 Dresden, Germany}
\author{E. \c{S}a\c{s}{\i}o\u{g}lu}
\author{C. Friedrich}
\author{S. Bl\"{u}gel}
\affiliation{Peter Gr\"{u}nberg Institut and Institute for
Advanced Simulation, Forschungszentrum J\"{u}lich and JARA,
52425 J\"{u}lich, Germany}

\begin{abstract}
We examine the mystery of the disputed high-magnetization
$\alpha^{\prime\prime}$-Fe$_{16}$N$_2$ phase, employing the
Heyd-Scuseria-Ernzerhof screened hybrid functional method,
perturbative many-body corrections through the GW approximation,
and onsite Coulomb correlations through the GGA+U method. We
present a first-principles computation of the
effective on-site Coulomb interaction (Hubbard $U$) between
localized 3$d$ electrons employing the constrained random-phase
approximation (cRPA), finding only somewhat stronger on-site
correlations than in bcc Fe. We find that the hybrid functional
method, the GW approximation, and the GGA+U method (using
parameters computed from cRPA) yield an average spin moment of
2.9, 2.6 -- 2.7, and 2.7 $\mu_B$ per Fe, respectively.
\end{abstract}

\maketitle \clearpage

\section{Introduction}

\begin{figure}[!htb]
\includegraphics[bb=14 14 256 249]{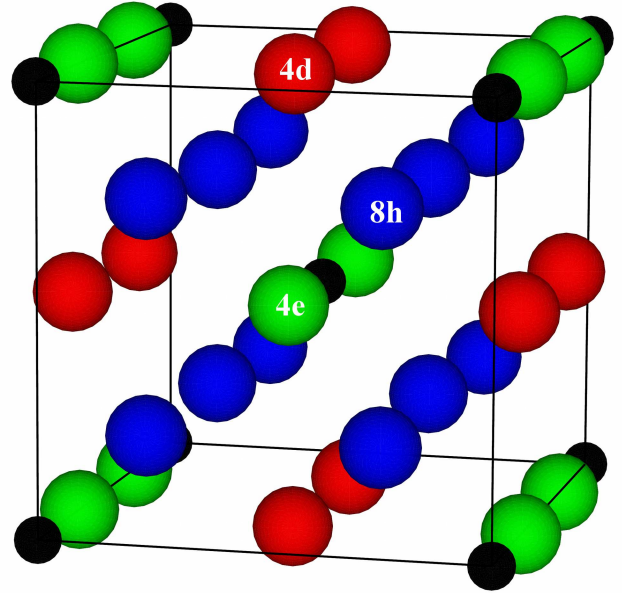}
\caption{(color online) Crystal structure of Fe$_{16}$N$_2$,
belonging to space group $I4/mmm$. We use the PBE-relaxed
structure in all calculations with $a=5.72$ \AA, $c=6.29$ \AA, $x
= 0.243$, and $z = 0.294$. We will frequently refer to the three
inequivalent Fe Wyckoff sites: $4d$ (red), $4e$ (green), and $8h$
(blue) (the N sites are black).}\label{fig:struct}
\end{figure}

Though discovered in 1951 by Jack,\cite{jack}
$\alpha^{\prime\prime}$-Fe$_{16}$N$_2$ (with crystal structure
pictured in Figure~\ref{fig:struct}) first drew the attention of
the magnetics community in 1972. It was then, 20 years later, that
Kim and Takahashi\cite{kt} reported polycrystalline, mixed-phase
Fe-N films with a saturation magnetization exceeding that of both
$\alpha$-Fe and Co$_{35}$Fe$_{65}$ ($\sim2\times10^{6}$ A/m).
However, it took another 20 years for the result to be reproduced
(and, in fact, surpassed) by Sugita {\it et
al.}\cite{sugita1,sugita2} Throughout the 1980s and '90s, other
measurements of Fe$_{16}$N$_2$ thin films were reported that
generally did not find this large magnetic
moment.\cite{kano,nakajima,mtaka1,mtaka2}

Concurrently, density-functional theory (DFT) electronic structure
calculations were
performed,\cite{coehoorn,coey,sakuma,ishida1,ishida2,min} finding
the moment per Fe ion to be modestly increased with respect to
bulk bcc Fe but far short of the 3.5 $\mu_B$ reported by Sugita
{\it et al.} It was shown\cite{lai} that LSDA+U\cite{lsdau}
calculations could yield an average moment comparable to that of
some experiments ($\sim 2.8 \mu_B$ per Fe), but the parameters ($U
\approx 3.94$, 1.0, and 1.34 eV on the $4d$, $4e$, and $8h$ sites,
respectively, with $J=U/10$) were obtained via an embedded-cluster
method with a small screening constant and were not calculated
from first principles. Additionally, the $J$ parameter is smaller
than usually considered appropriate for transition metals
(typically one chooses either an atomic-like $J$ of about $0.9$ eV
or else a more screened $J$ of about 0.6-0.7 eV).

Recently, further experimental evidence for the large
magnetization has arisen,\cite{ji_expt} as well as a companion
theoretical paper\cite{ji_theor} reporting enlarged Fe moments
achieved using LSDA+U (using $U=1.0$ eV for the $4d$ site, 4.0 eV
for the $4e$ and $8h$ sites, and $J$ = $U/10$). Ji {\it et al.}
motivate their parameters by proposing that the Fe sites in the
N-Fe octahedra form strongly correlated clusters in a metallic Fe
environment, choosing a small $U$ for the (within their model)
more metallic $4d$ sites and a large $U$ (chosen to be
intermediate between that of FeO and Fe) for the $4e$ and $8h$
sites. They suggest that this model is supported by XMCD spectra
that show additional features at the Fe sites not seen in bcc Fe
or other Fe-N phases.\cite{wang}

In the present work, we perform an extensive search for the
proposed large magnetization; we calculate the hyperfine field at
the three Fe sites and compare with published M\"ossbauer spectra;
we search for additional energy minima at moments away from the
theoretical prediction as a function of tetragonal distortion; we
apply the HSE06 hybrid-functional method\cite{hse06} and the GW
approximation\cite{gw} as implemented in \texttt{VASP}\cite{vasp}
to $\alpha^{\prime\prime}$-Fe$_{16}$N$_2$, testing the two methods
on bcc Fe to ensure that any enhancement of the moment we obtain
is genuine. Further, we compute the effective on-site Coulomb
interaction (Hubbard $U$) between localized 3$d$ electrons
employing the constrained random-phase approximation (cRPA)
\cite{cRPA,Sasioglu,cRPA_Sasioglu} (as implemented in the
\texttt{SPEX}\cite{Spex} extension of the
\texttt{FLEUR}\cite{Fleur} code), allowing us to provide for the
first time first-principles predictions for the $U$ and $J$
parameters. Finally, we present new PBE+U\cite{pbe} calculations
using these parameters and discuss their implications for existing
models.

\section{Computational Details}

The hyperfine field calculation and the study of the dependence of
the total energy on cell moment (fixed spin moment or FSM) and
tetragonal distortion was performed using the \texttt{FPLO}
code.\cite{Koepernik1999}We implemented the full relativistic
expression for the hyperfine field
\begin{eqnarray}
B_{HF} & = & \frac{\hbar^{2}}{ecm_{e}}\sum_{\nu}\left\langle\nu\left|
\underline{\boldsymbol{\alpha}}\cdot\left(\hat{\boldsymbol{\mu}}_{n}
\times\frac{\mathbf{r}}{r^{3}}\right)\right|\nu\right\rangle\label{eq:hf}
\end{eqnarray}
as given in Ref.~\onlinecite{Severin1997} and references therein
into \texttt{FPLO}\cite{Koepernik1999}. Note, that here we give
the general pre-factor to accomodate for proper units.
$|\nu\rangle$ are the solutions of the Kohn-Sham-Dirac equation,
while $\underline{\boldsymbol{\alpha}}=\left(\begin{array}{cc}
0 & \boldsymbol{\sigma}\\
\boldsymbol{\sigma} & 0
\end{array}\right)$, with $\boldsymbol{\sigma}$ being the vector of the
Pauli matrices. $\hat{\boldsymbol{\mu}}_{n}$ is the direction of
the nuclear spin moment. The wave functions $\Psi_{\nu}$ are
expanded in local atom centered orbitals in \texttt{FPLO}. The
effective integrand $\frac{1}{r^{2}}$ leads to a damping factor
for matrix elements between orbitals from different sites, which
allows to introduce the approximation that only terms with
orbitals belonging to the atom at which the nuclear spins sits
will be taken into account. The scalar relativistic hyperfine fields only contains the Fermi contact term, while
the full relativistic version contains all terms (including Fermi contact, orbital
and spin dipole-nuclear dipole), due to the intrinsic 4-component formulation of
Eq.~\eqref{eq:hf} and the use of 4-spinors. A non-relativistic
limit of this expression reveals all of the separate terms. The major
contributions come form the $s$-orbitals, for whom only the Fermi
contact term contributes. Although, the core states contribute a
large amount to the hyperfine field it has been
shown\cite{Ebert1996} that valence contributions can be sizable.
The core contribution depends on the spin polarization of the core wave functions
including the effects of the crystal exchange potential
and hence should be influenced and scaled by the local spin moment.
In \texttt{FPLO} the semi-core (Fe 3s,3p) states are treated like
valence states. For this reason we include the valence and
semi-core contributions via the onsite approximation as explained
above.

The FSM calculation was carried out within the PBE approximation
using a $8\times8\times8$ $k$-point mesh in a linear tetrahedron
method with Bl\"ochl corrections. Our \texttt{VASP} PBE+U (using
the fully-local double-counting term), HSE06 and GW calculations
used a plane-wave cut-off of 400 eV (29.4 Ry or 5.42
$a_{0}^{-1}$). The PBE+U (HSE06 and GW) calculations used an
$8\times8\times8$ ($6\times6\times6$) $\Gamma$-centered
Monkhorst-Pack $k$ mesh (also using the tetrahedron method with
Bl\"ochl corrections), employing a smaller $3\times3\times3$ mesh
for the exact-exchange sums. All of our \texttt{VASP} calculations
use the PAW\cite{paw} pseudopotentials of Kresse and
Joubert,\cite{kressepp} and all \texttt{VASP} moments are
calculated within a sphere of radius 1.3 \AA\ on the Fe sites.

To calculate the Hubbard $U$ parameter we employ the constrained
random-phase approximation (cRPA) \cite{cRPA} within the
full-potential linearized augmented-plane-wave (FLAPW) method
using maximally localized Wannier functions (MLWFs).
\cite{cRPA_Sasioglu,Max_Wan} The cRPA approach offers an efficient
way to calculate the effective Coulomb interaction $U$ and allows
to determine individual Coulomb matrix elements, e.g., on-site,
off-site, inter-orbital, intra-orbital, and exchange as well as
their frequency dependence. We use the  FLAPW method as
implemented in the \texttt{FLEUR} code \cite{Fleur} with the PBE
exchange-correlation potential\cite{pbe} for ground-state
calculations. A dense $16\times16\times16$ $\mathbf{k}$-point grid
is used. The MLWFs are constructed with the \texttt{Wannier90}
code \cite{Wannier90,Fleur_Wannier90}. The effective Coulomb
potential is calculated within the recently developed cRPA method
\cite{cRPA} implemented in the \texttt{SPEX} code \cite{Spex} (for
further technical details see Refs.\,\onlinecite{Sasioglu},
\onlinecite{cRPA_Sasioglu}, and \onlinecite{mpb}). We use a
$3\times3\times3$ $k$-point grid in the cRPA calculations.

In all calculations we use the PBE-relaxed structure with $a =
5.72$ \AA\ and  $c=6.29$ \AA\ (except in the FSM survey) and
internal parameters $x = 0.243$ and $z = 0.294$. As a final note,
we consider all employed electronic structure schemes
(\texttt{VASP}, \texttt{FLEUR}, \texttt{FPLO}) to be equivalent
with respect to numerical accuracy at the level required in the present study.
The use of three different packages is motivated by the different
implementations available in these codes.

\section{Results and Discussion}

\subsection{Hyperfine Field}

The hyperfine field provides a picture of the local magnetic
structure that, unlike measurements of the saturation
magnetization, does not require accurate estimation of the volume
of a sample or its component phases. M\"ossbauer spectroscopy has
been performed in many previous
works\cite{moriya,sugita1,sugita2,mtaka2,ortiz,nakajima2,okamoto},
and the hyperfine field has been
calculated\cite{coehoorn,ishida1,ishida2,coey2} from DFT. Our
calculated $B_{hf}$ (found along with our calculated Fe moments in
Table~\ref{tab:bhf}) agrees well with these past results; we find
that the Fe sites with N nearest-neighbors exhibit approximately
the same field (-23 and -22 T on the $4e$ and $8h$ sites), while
$B_{hf} = -31$ T for the $4d$ sites. If we note, as previous
authors have,\cite{coehoorn} that DFT underestimates the hyperfine
field by a substantial, though nearly static, amount ($\sim 8$ T
in this case), then we also find reasonable agreement with some of
the experimental reports. Particularly, we agree well with Refs.
\onlinecite{mtaka2}, \onlinecite{moriya}, and  \onlinecite{ortiz}.
Although our hyperfine fields agree numerically with those of
Moriya {\it et al.},\cite{moriya} they claimed that the largest
hyperfine field was to be found in the $8h$ site, a claim that is
difficult to reconcile with the predicted relative magnitudes of
the moments and the similar environment of the $4e$ and $8h$
sites. We note, however, that this assignment of the hyperfine
fields agrees better with the moments in the recent LSDA+U study
of Ji {\it et al.}.\cite{ji_theor} We cannot offer any new
explanation for Sugita {\it et al.}'s larger 46 T
field\cite{sugita1} nor the presence of only one M\"ossbauer
sextet in their later single-phase sample.\cite{sugita2}

\begin{table}[htb]
\begin{ruledtabular}
\begin{tabular}{lccc}
Site & Spin Moment ($\mu_B$) & $B_{hf}$ (T) (scal.-rel.) & $B_{hf}$ (T) (rel.) \\
\hline
$4d$ & 2.85 & -34 & -31 \\
$4e$ & 2.17 & -25 & -23 \\
$8h$ & 2.36 & -25 & -22 \\
\end{tabular}
\end{ruledtabular}
\caption{Spin moments calculated within scalar-relativistic PBE
and hyperfine fields  (both fully- and scalar-relativistic) for
each Fe site calculated within PBE using the \texttt{FPLO}
code.}\label{tab:bhf}
\end{table}

\subsection{Fixed Spin-Moment Survey}

Generally, expansion of the lattice may not be an efficient  means
of increasing the magnetization of a material, as the enhancement
of the spin moments may not outpace the increase in volume.
However, it is known that fcc Fe, while ordinarily nonmagnetic,
enters a high spin state upon expansion of the cell
volume.\cite{fccFe} Therefore, we have explored the energy
landscape as a function of total (spin) cell moment and
$\frac{c}{a}$, allowing the former to range from 34 -- 48 $\mu_B$
(corresponding to average spin moments of 2.12 -- 3.0 $\mu_B$ per
Fe) and the latter from 1.0 -- 1.5 (holding $a$ fixed in one set
of calculations and volume fixed in another). We only constrain
the total spin moment of the cell and not the magnitude of the
individual moments. In principle, the moments of the three
inequivalent Fe sites could be arranged in many ways to obtain the
same total spin moment; however, we simply accept the converged
result for each structure and total moment without seeking out
other possible minima.

\begin{figure}[htb]
\includegraphics[bb= 14 14 256 355]{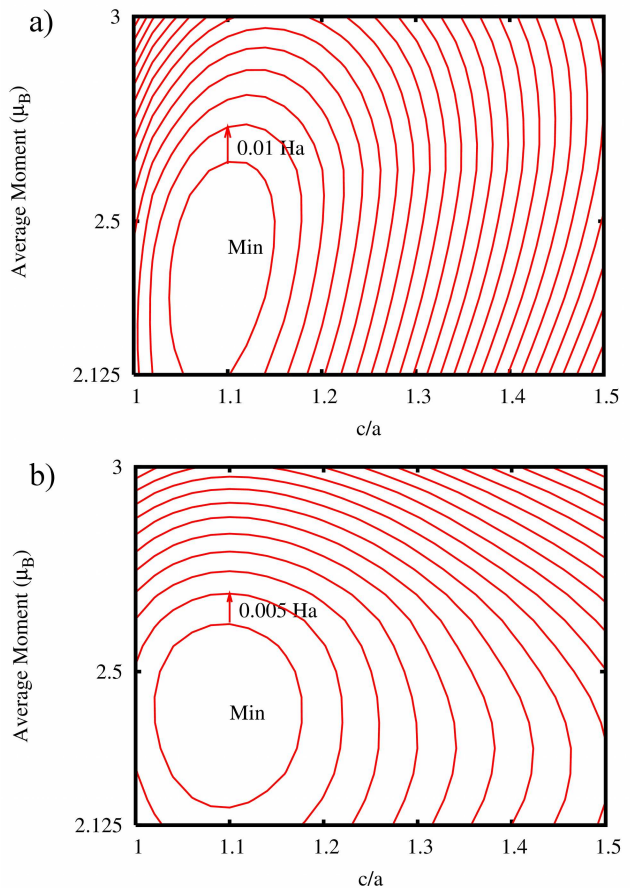}
\caption{(color online) Energy landscape of Fe$_{16}$N$_2$ as a
function of tetragonal distortion and average moment per Fe with
a) $a$ held fixed and b) volume held fixed. Each contour
represents an increase of a) 10 mHa (0.27 eV) or b) 5 mHa (0.14
eV). There are no additional local energy minima in the parameter
space examined. For fixed $a$, $\frac{c}{a}=1.5$ gives $\mu_{avg}
= 2.81 \mu_B$. Although the average moment is higher at this
point, it does not overcome the increase in volume, and the
magnetization is only 84\% of the magnetization of the PBE
structure. For fixed volume, the average moment remains close to
the PBE value, decreasing slightly as $c$ increases above $\sim
1.1a$.}\label{fig:fsm}
\end{figure}

The results may be seen in Figure~\ref{fig:fsm}. We note that no
additional local energy minima were observed apart from the
PBE-relaxed structure ($a = 5.72$\ \AA, $c = 6.29$\ \AA, $c/a \approx 1.10$)
and moment (2.44 $\mu_B$). Although the energy minimum does
tend to shift to higher moments with the increase of the volume
through $\frac{c}{a}$, the enhancement is not sufficient to
produce an increase in the magnetization. With $a$ held fixed at
$a_{\text{expt}}=5.72$ \AA, the average spin moment per Fe reaches
2.81 $\mu_B$ at $\frac{c}{a} = 1.5$, giving a magnetization of
$1.49\times10^6$ A/m, compared to $1.77\times10^6$ A/m at the
experimental $\frac{c}{a}$ and $1.75\times10^6$ A/m in bcc Fe. If
the volume is held fixed, the average moment does not depend
strongly on $\frac{c}{a}$, remaining close to the PBE value
throughout and decreasing to about $2.25 \mu_B$ at $\frac{c}{a} =
1.5$. This supports the standard understanding of the LSDA- or
GGA-predicted increase in the moment as arising from increased
cell volume.

\subsection{HSE06 and GW}

It is possible that DFT cannot fully account for the physics that
would give rise to greatly enhanced magnetization in
$\alpha^{\prime\prime}$-Fe$_{16}$N$_2$, so we have also considered
methods that have arisen since the last wave of theoretical
investigation into this material subsided. The HSE06 screened
hybrid functional method entails only a moderate increase in
computational time with respect to PBE, and the inclusion of a
static screening parameter for the exact exchange term allows for
the treatment of metallic systems---unlike the parent Hartree-Fock
method---as well as speeding up calculation further. HSE06 follows
PBE0 in its formulation of the exchange-correlation energy, given
by
\begin{equation}
E_{xc} = \frac{1}{4}E_{x}^{HF,SR} + \frac{3}{4}E_{x}^{PBE,SR} +
E_{x}^{PBE,LR} + E_{c}^{PBE}
\end{equation}
The aforementioned screening parameter $\mu = 0.2$ \AA$^{-1}$\
partitions the exchange term into a short-range and a long-range
component, achieved by appending erfc$(\mu r)$ (the complementary
error function) to the short-range terms and erf$(\mu r)$ to the
long-range term.\cite{hse06}

The GW approximation improves upon Hartree-Fock by treating
electrons as  dressed quasiparticles interacting via a screened
Coulomb operator $W$. This replaces the purely real
exchange-correlation potential with a complex self energy
$\Sigma= -iGW$.
In the initial step, the Green's function $G$ and the screened
Coulomb operator $W$ are calculated from the wave functions
obtained from a converged DFT calculation. The computation of $W$
via the RPA is time-consuming, and consequently some shortcuts are
sometimes employed. So-called ``one-shot'' GW or G$_0$W$_0$ is
performed by calculating the quasiparticle energies using only
these initial quantities and yields improved results compared to
LSDA.\cite{faleev,vsg} Nevertheless, the ``one-shot'' method still
underestimates band-gaps due to the inaccuracies inherent in using
an LSDA-obtained $W$, and improvement can be obtained by iterating
$G$ and $W$ to self-consistency. We present results from
G$_0$W$_0$, GW$_0$, and GW in this work.

\begin{figure*}[!tb]
\includegraphics[bb=14 14 507 596]{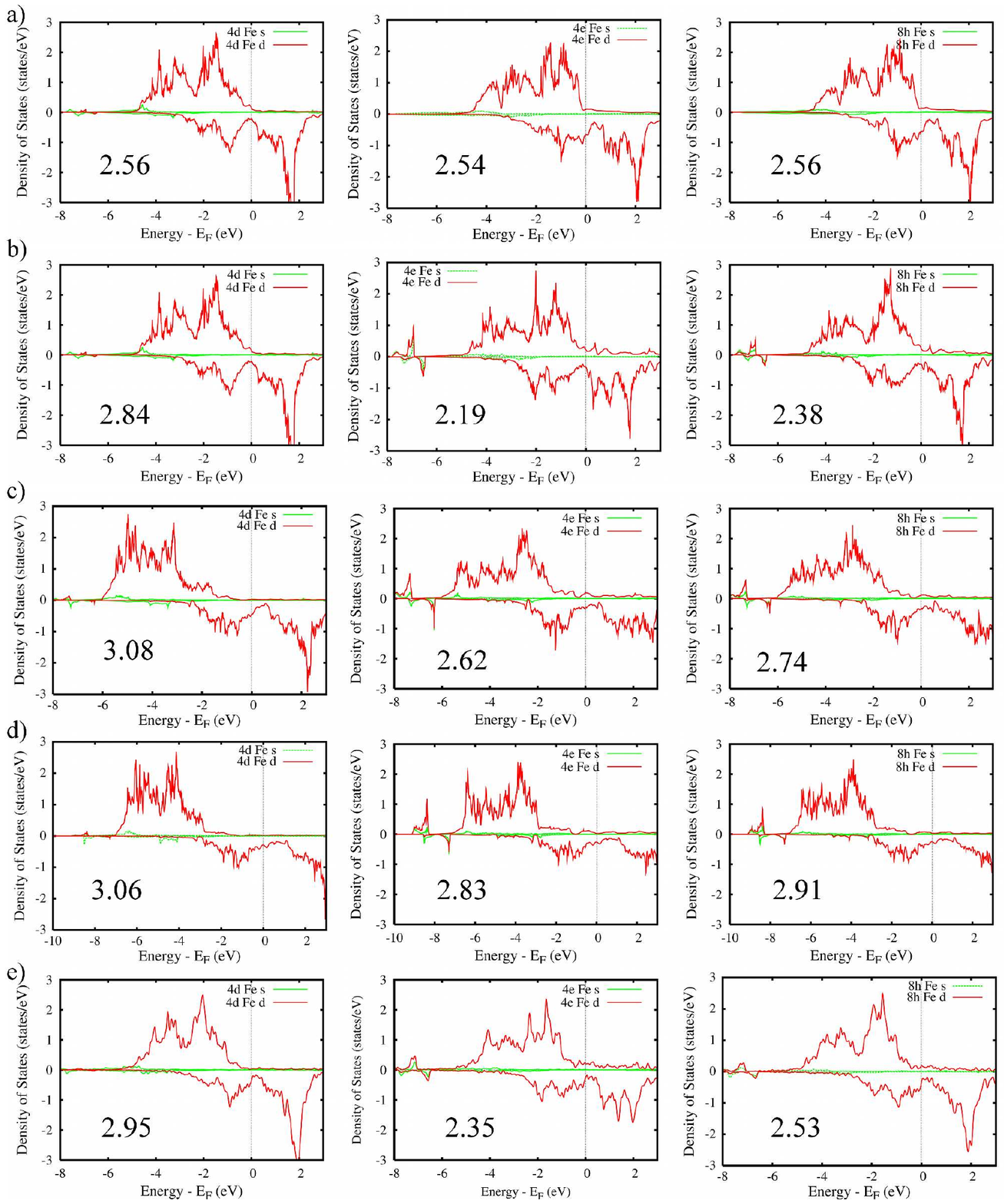}
\caption{(color online) $s$ and $d$ Partial density of states and
spin moments (in $\mu_B$) for the $4d$, $4e$, and $8h$ Fe sites
(left, middle and right columns) for a) a fictitious
Fe$_{16}$N$_0$ structure (within PBE) in which the Fe atoms retain
their $\alpha^{\prime\prime}$ positions. Here, we see that the
majority channel is already nearly fully occupied due to the
volume expansion induced by the N atoms. (b) PBE pDOS showing the
effect of Fe -- N hybridization: increasing the moment of the
second-neighbor $4d$ site at the expense of the $4e$ and $8h$
sites. (c-e) PBE+U, HSE06, and GW$_0$, respectively (the reader
should note the difference in the scale of the $x$ axis in the HSE
plots): PBE+U and HSE06 each give large moments at each site, but
predict an average moment only slightly larger than in bcc Fe (for
which they also give very large moments). The pDOS for all
\texttt{VASP}-GW methods considered in the text are in qualitative agreement
with those displayed.}\label{fig:alldos}
\end{figure*}

Figure~\ref{fig:alldos} shows the partial density of states (pDOS)
of each Fe site in HSE06 and GW. For comparison, we include the
PBE-calculated pDOS for Fe$_{16}$N$_2$ and a fictitious
``Fe$_{16}$N$_0$'' structure obtained by removing the N atoms
without relaxing the structure. This latter case shows that,
within PBE, Fe approaches the strong ferromagnetic state, with the
majority $d$ states nearly fully occupied, upon the N-induced
volume expansion, yielding an average moment of 2.56 $\mu_B$ per
Fe and a magnetization of $1.84\times10^{6}$ A/m, about a 5\% increase
over bcc Fe (with a bulk magnetization of $1.75\times10^6$ A/m).
The HSE06 pDOS shows a greatly enhanced exchange
splitting with respect to PBE-Fe$_{16}$N$_0$ and
GW-Fe$_{16}$N$_2$, leading to an average moment of 2.86 $\mu_B$
per Fe ($M = 2.06\times10^6$ A/m), whereas GW yields a more
moderate 2.57-2.70 $\mu_B$ per Fe ($M = 1.85$--$1.95\times10^6$ A/m). The
calculated spin moment at each site can be found in
Table~\ref{tab:mom}.

\begin{table*}[!tb]
\begin{ruledtabular}
\begin{tabular}{lccccc}
Method & $4d$ Site ($\mu_B$) & $4e$ Site ($\mu_B$) & $8h$ Site ($\mu_B$) & Average ($\mu_B$) & bcc Fe ($\mu_B$) \\
\hline
PBE & 2.84 & 2.19 & 2.38 & 2.44 & 2.23 \\
PBE+U & 3.08 & 2.62 & 2.74 & 2.71 & 2.67 \\
HSE06 & 3.06 & 2.83 & 2.91 & 2.86 & 2.85 \\
G$_0$W$_0$ & 2.90 & 2.31 & 2.49 & 2.57 & 2.33 \\
GW$_0$ & 2.95 & 2.35 & 2.53 & 2.64 & 2.62 \\
GW & 2.96 & 2.41 & 2.57 & 2.66 & 2.65 \\
GW ($s,p$ val.) & 3.00 & 2.50 & 2.64 & 2.70 & 2.59 \\
\end{tabular}
\end{ruledtabular}
\caption{Calculated spin moments for all methods presented in this
work. The PBE+U results were obtained using the cRPA-obtained
interaction parameters $U=3.99$, 3.12, and 3.52 eV for the $4d$,
$4e$, and $8h$ sites, respectively ($J=0.64$, 0.59, and 0.61 eV),
except for bcc Fe, for which we used $U=3.16$ eV and $J=0.68$ eV
as in Ref.~\onlinecite{cRPA_Sasioglu}. GW ($s,p$ val) denotes a
\texttt{VASP}-GW calculation in which the Fe $3s$ and $3p$
electrons are treated on the same level as the $3d$ and $4s$.
}\label{tab:mom}
\end{table*}

In the absence of experimental photo- or x-ray-emission data to
which  to compare, we must test the validity of the calculated
moments by calculating the moments of better established
materials. The last column in this table shows the calculated spin
moment for bcc Fe from PBE, PBE+U (which will be discussed in
detail in the following section), HSE06, and GW. Our HSE06 result
for bcc Fe agrees with previous work\cite{jang} and demonstrates
that, although the screened hybrid functional method improves on
the Hartree-Fock treatment of metallic systems, it can
overestimate the strength of the exchange and yield un-physical
high spin states. However, we also note that the calculated bcc Fe
spin moment is not necessarily directly proportional to the
calculated moments in $\alpha^{\prime\prime}$-Fe$_{16}$N$_2$, so
it is possible that the bcc Fe moment does not completely
determine the accuracy of a method in this case.

\subsection{cRPA and PBE+U}

Previous attempts to explain the experiments that find
high-magnetization have turned to LDA+U to describe the
correlation effects that may be present in
$\alpha^{\prime\prime}$-Fe$_{16}$N$_2$. However, as no
first-principles calculations of the interaction parameters
existed, it was necessary to motivate the choice of $U$ (and $J$)
by analogy with other systems or by applying a model. In
particular, the explanation for the enhanced magnetization
proposed by Ji {\it et al.}\cite{ji_expt,ji_theor} and Wang {\it
et al.}\cite{wang} requires that the Fe sites with N
nearest-neighbors be more strongly-correlated than the $4d$ sites,
which have no N neighbors. Without a set of firmly-established
parameters, it is difficult to progress in understanding this
system, as the calculated moment is directly dependent on $U$ and
$J$ (see, e.g. Figure 3 in Ref. \onlinecite{ji_theor}).

Recently, the cRPA has been proposed as a first-principles method
of obtaining the screened Coulomb matrix within a Wannier
basis.\cite{cRPA,Sasioglu,cRPA_Sasioglu} Within the RPA, the
polarizability $P$ can be written
\begin{align}
P(\mathbf{r},\mathbf{r}^\prime,\omega) = \sum_\sigma\sum_{n}^{\text{occ}} \sum_{m}^{\text{unocc}}
\left[\frac{\psi^{*}_{\sigma{}n}(\mathbf{r})\psi_{\sigma{}m}(\mathbf{r})\psi^{*}_{\sigma{}m}(\mathbf{r}^\prime)
\psi_{\sigma{}n}(\mathbf{r}^\prime)}{\omega-\varepsilon_{\sigma{}m} + \varepsilon_{\sigma{}n} + i\delta} \right.\nonumber \\
\left. - \frac{\psi_{\sigma{}n}(\mathbf{r})\psi^{*}_{\sigma{}m}(\mathbf{r})\psi_{\sigma{}m}(\mathbf{r}^\prime)\psi^{*}_{\sigma{}n}
(\mathbf{r}^\prime)}{\omega+\varepsilon_{\sigma{}m} - \varepsilon_{\sigma{}n} - i\delta}\right],
\end{align}
where the $\psi_i$ and $\varepsilon_i$ are the DFT wave functions
and  their eigenvalues, and $\sigma$ runs over both spin channels.
If one separates $P$ into $P_d$,
containing the correlated orbitals, and $P_r$, containing the
rest, and if one considers the unscreened Coulomb operator $v$,
one can write\cite{cRPA,cRPA_Sasioglu}
\begin{align}
U =[1-vP_r]^{-1}v \\
\tilde{U} = [1-UP_d]^{-1}U
\end{align}
The matrix elements of the effective Coulomb potential $U$ in the
MLWF basis are given by
\begin{align}
U_{\mathbf{R}n_1 n_3;n_4 n_2}(\omega) = \iint
w_{n_1\mathbf{R}}^{*}(\mathbf{r}) w_{n_3\mathbf{R}}
(\mathbf{r}) U(\mathbf{r},\mathbf{r}^{\prime};\omega) \nonumber\\
\times w_{n_4\mathbf{R}}^{*} (\mathbf{r}^{\prime})w_{n_2\mathbf{R}}
(\mathbf{r}^{\prime})\:d^3r\:d^3r^{\prime},
\end{align}
where $w_{n\mathbf{R}}(\mathbf{r})$ is the MLWF at site
$\mathbf{R}$  with orbital index $n$ and $U(\mathbf{r},
\mathbf{r}^{\prime};\omega)$ is calculated within the cRPA. Strictly speaking, the Wannier
functions are spin dependent. However, we find that this spin dependence
affects the values only little. For simplicity, we ignore the spin
dependence here and give the spin-averaged values in the following.

In our Spex-cRPA calculation, we choose the Fe $d$ orbitals as our
correlated subspace and compute the interaction parameters found
in Table\,\ref{tab:crpa}. Quantities with tildes are obtained from
the fully screened Coulomb matrix $\tilde{U}$, while plain symbols
are the $sp$-screened quantities that enter into the PBE+U
calculations. The $U$, $U^\prime$, and $J$ (and their
fully-screened counterparts) are averaged at each site as follows:
\refstepcounter{equation}
\begin{align}
U_{DFT+U} = F^0 = \frac{1}{25}\sum_{m,n} U_{mnmn} \tag{\theequation{}a} \\
U = \frac{1}{5}\sum_m U_{mmmm} \tag{\theequation{}b} \\
U^\prime = \frac{1}{10}\sum_{m<n} U_{mnmn} \tag{\theequation{}c} \\
J = \frac{1}{10}\sum_{m<n} U_{mnnm}  \tag{\theequation{}d}
\end{align}

\begin{table}[!htb]
\begin{ruledtabular}
\begin{tabular}{lccccccc}
Site & $U_{DFT+U}$ & $U$ & $U^\prime$ & $J$ & $\tilde{U}$ & $\tilde{U}^\prime$ & $\tilde{J}$  \\
\hline
$4d$ & 3.99 & 5.02 & 3.74 & 0.64 & 1.80 & 0.71 & 0.53 \\
$4e$ & 3.12 & 4.14 & 2.95 & 0.59 & 1.56 & 0.55 & 0.49 \\
$8h$ & 3.52 & 4.50 & 3.27 & 0.61 & 1.68 & 0.62 & 0.51 \\
\end{tabular}
\end{ruledtabular}
\caption{The calculated on-site interaction parameters (all in eV)
from cRPA for $\alpha^{\prime\prime}$-Fe$_{16}$N$_2$, showing a
small increase in correlation with respect to bcc Fe. Quantities
with a tilde are computed from the fully-screened Coulomb
potential, while plain quantities are computed from the partially
screened potential (omitting $d-d$ screening). $U_{DFT+U}$ is the
$U$ parameter that enters into the PBE+U
calculations.}\label{tab:crpa}
\end{table}

We note that these parameters differ both quantitatively and
qualitatively from previously proposed models, particularly those
that suggest large differences in correlation strength between Fe
sites. The spin moments from PBE+U, for Fe$_{16}$N$_2$ as well as
bcc Fe, can be found in Table~\ref{tab:mom}. The PBE+U spin moment
for bcc Fe was calculated using the interaction parameters
computed in Ref.~\onlinecite{cRPA_Sasioglu}---$U = 3.16$ eV and $J
= 0.68$ eV. We use the fully-local (FLL) double counting correction in the calculation of
both the bcc Fe and the Fe$_{16}$N$_2$ moments. Although this choice may
seem strange in metallic systems, the around-mean-field (AMF) term opposes
the formation of moments in general\cite{Ylvisaker2009} and here
produces moments $\sim 1 \mu_B$ below the expected value in bcc Fe.
It should be noted, however,  that the choice of the double counting term
in PBE + U is not unique and thus leaves an ambiguity in the calculated
moments even if $U$ and $J$ were computed with a well-defined method..

\subsection{Orbital Moment}

In solids, the orbital moment is typically nearly quenched, but in
some  extreme cases, such as UN,\cite{brooks} the orbital moment
can be comparable to the spin moment. PBE calculations give an
orbital moment per Fe of only 0.05 $\mu_B$ in bcc Fe
(Table~\ref{tab:orb}), but this may be increased somewhat in
Fe$_{16}$N$_2$. To explore this possibility, we calculated the
orbital moment within PBE, PBE+orbital polarization correction
(OPC),\cite{nordstrom} PBE+U (using the cRPA parameters), and
``one-shot'' G$_0$W$_0$ using \texttt{FPLO} (for the OPC
calculation) and \texttt{VASP} (for the rest). Each method shows a
small increase in orbital moment compared to bcc Fe, yielding
about 0.1 -- 0.2 $\mu_B$ per Fe atom and an increase of 0.01 --
0.05 $\mu_B$ over bcc Fe. This small increase cannot explain
those results that claim average Fe moments in excess of
3 $\mu_B$. Our PBE+U and G$_0$W$_0$ results predict average
total (spin + orbital) moments of 2.88 and 2.63 $\mu_B$, respectively.

\begin{table*}[!htb]
\begin{ruledtabular}
\begin{tabular}{lccccc}
Method & $4d$ Site ($\mu_B$) & $4e$ Site ($\mu_B$) & $8h$ Site ($\mu_B$) & Average ($\mu_B$) & bcc Fe ($\mu_B$) \\
\hline
PBE & 0.06 & 0.06 & 0.05 & 0.06 & 0.05 \\
PBE+U & 0.20 & 0.16 & 0.16 & 0.17 &  0.12 \\
PBE+OPC & 0.09 & 0.11 & 0.10 & 0.10 & 0.09 \\
G$_0$W$_0$ & 0.06 & 0.06 & 0.05 & 0.06 & 0.05 \\
\end{tabular}
\end{ruledtabular}
\caption{Calculated orbital moments in Fe$_{16}$N$_2$ within PBE,
PBE+U, PBE+OPC, and G$_0$W$_0$. The orbital moment is increased by
only 0.01 - 0.03 $\mu_B$ per Fe with respect to bcc
Fe.}\label{tab:orb}
\end{table*}

\section{Summary}

We have examined the electronic and magnetic structure of
$\alpha^{\prime\prime}$-Fe$_{16}$N$_2$ within PBE, PBE+U, HSE06,
and GW. Within PBE, we find spin moments and hyperfine fields that
agree with past results, and we do not find that any
high-magnetization state arises as $\frac{c}{a}$ changes from the
experimental value. We have provided effective Coulomb interaction parameters calculated via
cRPA and have used them in our PBE+U calculations. We find that
PBE+U and HSE06 gives average spin moments per Fe of 2.71 and 2.86
$\mu_B$ but also greatly overestimate the moment of bcc Fe (experimentally
 about 2.2 $\mu_B$). GW
gives smaller moments, 2.57 - 2.70 $\mu_B$ per Fe, a slight
increase over the PBE moment. G$_0$W$_0$, GW$_0$, and GW all
overestimate the bcc Fe spin moment by different amounts despite
their similar predictions for Fe$_{16}$N$_2$, with G$_0$W$_0$ giving
the most reasonable bcc Fe moment due to its close dependence on
the PBE result. In all cases, we
find that the $4e$ and $8h$ sites have smaller moments than that
on the $4d$ site.

We have also presented calculations of the orbital moment on the
Fe sites obtained within PBE, PBE+OPC, PBE+U, and G$_0$W$_0$. We
find that the orbital moment is not completely quenched and may
add 0.1 -- 0.2 $\mu_B$ to the average total moment per Fe, a small
increase over bcc Fe.

In order to evaluate the varying results found above, one must
understand the purposes of and approximations inherent in the
methods presented. In addition to the shortcomings of the mean-field-like treatment
of correlations within PBE+U, there are two notable avenues for error in this method: the need
to choose the $U$ and $J$ parameters and the lack of \emph{a priori} justification
for the double-counting corrections. Dependence on the choice of interaction
parameters is not a fundamental problem and can be alleviated as we have done
here by computing them through some appropriate first-principles method. The
choice between the FLL or AMF double-counting corrections, while straightforward
when treating insulators, can be less obvious in semi-localized magnetic systems,
and furthermore no method exists for determining the exact form of the
correction. The hybrid functional method's dependence on parameters is
fundamental to the approach, although it is mitigated somewhat by the use
of predetermined parameters such as in HSE06. However, these parameters
were primarily chosen to produce reasonable band gaps and may need to
be altered to properly treat metallic systems. In principle, the GW approximation
should be the most accurate of those presented here. The G$_0$W$_0$
and GW$_0$ methods maintain good contact with the PBE results while
incorporating first-order exchange and correlation effects. However,
some care must still be taken; we have shown that the results do depend on
which electrons are treated as valence and which are absorbed into the core
pseudopotential. Lastly, we note the need for additional, repeatable
experiments that probe the electronic structure of the material in
order to provide a better basis for comparison with theory.

\section{Acknowledgments}
H.S. and W.H.B. acknowledge the support of NSF MRSEC Grant No.
DMR-0213985 and the use of computing resources from the Alabama
Supercomputer Center. M.R. would like to thank Joachim Wecker and
Manfred R\"uhrig for discussion. E.\c{S}, C.F and S.B. acknowledge
the support of DFG through the Research Unit FOR-1346.


\begin{thebibliography}{00}

\bibitem{jack}
K. H. Jack, Proc. R. Soc. A {\bf 208}, 200 (1951).


\bibitem{kt}
T. K. Kim and M. Takahashi, Appl. Phys. Lett. {\bf 20}, 492
(1972).


\bibitem{sugita1}
Y. Sugita, K. Mitsuoka, M. Komuro, H. Hoshiya, Y. Kozono, and M.
Hanazono, J. Appl. Phys. {\bf 70}, 5977 (1991).


\bibitem{sugita2}
Y. Sugita, H. Takahashi, M. Komuro, K. Mitsuoka, and A. Sakuma, J.
Appl. Phys. {\bf 76}, 6637 (1994).


\bibitem{kano}
A. Kano, N. S. Kazama, H. Fujimori, and T. Takahashi, J. Appl.
Phys. {\bf 53}, 8332 (1982).


\bibitem{nakajima}
K. Nakajima and S. Okamoto, Appl. Phys. Lett. {\bf 56}, 92 (1990).


\bibitem{mtaka1}
M. Takahashi, H. Shoji, H. Takahashi, T. Wakiyama, M. Kinoshita,
and W. Ohta, IEEE Trans. Magn. {\bf 29}, 3040 (1993).


\bibitem{mtaka2}
M. Takahashi, H. Takahashi, H. Nashi, H. Shoji, T. Wakiyama, and
M. Kuwabara, J. Appl. Phys. {\bf 79}, 5564 (1996).


\bibitem{coehoorn}
R. Coehoorn, G. H. O. Daalderop, and H. J. F. Jansen, Phys. Rev. B
{\bf 48}, 3830 (1993).


\bibitem{coey}
J. M. D. Coey, J. Appl. Phys. {\bf 76}, 6632 (1994).


\bibitem{sakuma}
A. Sakuma, J. Magn. Magn. Mater. {\bf 102}, 127 (1991);  A.
Sakuma, J. Phys. Soc. Jpn. {\bf 60}, 2007 (1991); A. Sakuma, J.
Phys. Soc. Jpn. {\bf 61}, 223 (1992).


\bibitem{ishida1}
S. Ishida and K. Kitawatase, J. Mag. Magn. Mat. {\bf 104-107},
1933 (1992).


\bibitem{ishida2}
S. Ishida, K. Kitawatase, S. Fujii, and S. Asano, J. Phys.
Condens. Matter {\bf 4}, 765 (1992).


\bibitem{min}
B. I. Min, Phys. Rev. B {\bf 46}, 8232 (1992).


\bibitem{lai}
W. Y. Lai, Q. Q. Zheng, and W. Y. Hu, J. Phys. Cond. Mat. {\bf 6},
L259 (1994).


\bibitem{ji_expt}
N. Ji, L. F. Allard, E. Lara-Curzio, and J.-P. Wang, Appl. Phys.
Lett. {\bf 98}, 092506 (2011).


\bibitem{ji_theor}
N. Ji, X. Liu, and J.-P. Wang, New Journal of Physics {\bf 12},
063032 (2010).


\bibitem{lsdau}
V .I. Anisimov, J. Zaanen , and O.K. Andersen Phys. Rev. B {\bf
44}, 943 (1991).



\bibitem{wang}
J.-P. Wang, N. Ji, X. Liu, Y. Xu, C. S\'{a}nchez-Hanke, Y. Wu, F.
M. F. de Groot, L. F. Allard, and E. Lara-Curzio, IEEE Trans.
Magn. {\bf 48}, 1710 (2012).


\bibitem{hse06}
J. Heyd, G. E. Scuseria, and M. Ernzerhof, J. Chem. Phys. {\bf
124}, 219906 (2006).


\bibitem{gw}
L. Hedin, Phys. Rev. {\bf 139}, A796 (1965).


\bibitem{vasp}
G. Kresse and J. Furthm\"{u}ller, Comput. Mater. Sci. {\bf 6}, 15
(1996).



\bibitem{cRPA}
F. Aryasetiawan, M. Imada, A. Georges, G. Kotliar, S. Biermann, and A.
I. Lichtenstein, Phys. Rev. B \textbf{70}, 195104 (2004); F.
Aryasetiawan, K. Karlsson, O. Jepsen, and U. Sch\"{o}nberger,
Phys. Rev. B \textbf{74}, 125106 (2006); T. Miyake, F.
Aryasetiawan, and M. Imada Phys. Rev. B \textbf{80}, 155134
(2009); B-C. Shih, Y. Zhang, W. Zhang, and P. Zhang, Phys. Rev. B
\textbf{85}, 045132 (2012); T. O. Wehling, E.
\c{S}a\c{s}{\i}o\u{g}lu, C. Friedrich, A. I. Lichtenstein, M. I.
Katsnelson, and S. Bl\"{u}gel, Phys. Rev. Lett. \textbf{106},
236805 (2011).


\bibitem{Sasioglu}
E. \c{S}a\c{s}{\i}o\u{g}lu, A. Schindlmayr, C. Friedrich, F.
Freimuth and S. Bl\"{u}gel, Phys. Rev. B. \textbf{81}, 054434
(2010).


\bibitem{cRPA_Sasioglu}
E. \c{S}a\c{s}{\i}o\u{g}lu, C. Friedrich, and S. Bl\"{u}gel, Phys.
Rev. B \textbf{83}, 121101(R) (2011).


\bibitem{Spex}
C. Friedrich, S. Bl\"{u}gel and A. Schindlmayr, Phys. Rev. B.
\textbf{81}, 125102 (2010).


\bibitem{Fleur}
http://www.flapw.de



\bibitem{pbe}
J. P. Perdew, K. Burke, and M. Ernzerhof, Phys. Rev. Lett. {\bf
77}, 3865 (1996).



\bibitem{Severin1997}
L. Severin and M. Richter and L. Steinbeck, Phys. Rev. B {\bf 55},
9211 (1997).


\bibitem{Koepernik1999}
K. Koepernik and H. Eschrig, Phys. Rev. B {\bf 59}, 1743 (1999),
http://www.fplo.de.



\bibitem{Ebert1996}
M. Battocletti and H. Ebert and H. Akai, Phys. Rev. B {\bf 53},
9776 (1996).


\bibitem{paw}
P. E. Bl\"{o}chl, Phys. Rev. B {\bf 50}, 17953 (1994).


\bibitem{kressepp}
G. Kresse and D. Joubert, Phys. Rev. B {\bf 59}, 1758 (1999).


\bibitem{Max_Wan}
N. Marzari and D. Vanderbilt, Phys. Rev. B \textbf{56}, 12847
(1997).

\bibitem{Fleur_Wannier90}
F. Freimuth, Y. Mokrousov, D. Wortmann, S. Heinze, and S.
Bl\"{u}gel, Phys. Rev. B \textbf{78}, 035120 (2008).


\bibitem{Wannier90}
A. A. Mostofi, J. R. Yates, Y.-S. Lee, I. Souza, D. Vanderbilt,
and N. Marzari, Comput. Phys. Commun. \textbf{178}, 685 (2008).


\bibitem{mpb}
C. Friedrich, A. Schindlmayr, and S. Bl\"{u}gel, Comp. Phys. Comm.
{\bf 180}, 347 (2009).


\bibitem{moriya}
T. Moriya, Y. Sumitomo, H. Ino, and F. E. Fujita, J. Phys. Soc.
Jpn. {\bf 35}, 1378 (1973).


\bibitem{ortiz}
C. Ortiz, G. Dumpich, A. H. Morrish, Appl. Phys. Lett. {\bf 65},
2737 (1994).


\bibitem{okamoto}
S. Okamoto, O. Kitakami, Y. Shimada, J. Mag. Magn. Mat. {\bf 208},
102 2000.


\bibitem{nakajima2}
K. Nakajima, T. Yamashita, M. Takata, and S. Okamoto, J. Appl.
Phys. {\bf 70}, 6033 (1991).


\bibitem{coey2}
J M D. Coey, K. O’Donnell, Q. Qi, E. Touchais, and J. H. Jack,
J. Phys. Cond. Mat. {\bf 6}, L23 (1994).


\bibitem{fccFe}
V. L. Moruzzi, P. M. Marcus, K. Schwarz, and P. Mohn, Phys. Rev. B
{\bf 34}, 1784 (1986).


\bibitem{vsg}
M. van Schilfgaarde, T. Kotani, and S. Faleev, Phys. Rev. Lett.
{\bf 96}, 226402 (2006).


\bibitem{faleev}
S. V. Faleev, M. van Schilfgaarde, and T. Kotani, Phys. Rev. Lett.
{\bf 93}, 126406 (2004).


\bibitem{jang}
Y.-R. Jang and B. D. Yu, J. Mag. {\bf 16}, 201 (2011).


\bibitem{brooks}
M. S. S. Brooks and P. J. Kelly, Phys. Rev. Lett. {\bf 51}, 1708
(1983).


\bibitem{Ylvisaker2009}  E. R. Ylvisaker, W. E. Pickett, and K. Koepernik, Phys. Rev. B {\bf 79}, 035103 (2009).


\bibitem{nordstrom}
L. Nordstr\"om, M. S. S. Brooks, and B. Johansson, J. Phys.: Cond.
Matt. {\bf 4}, 3261 (1992).

\end{thebibliography}
\end{document}